\documentclass[aps,prb,reprint,epsfig,amssymb,superscriptaddress,showpacs]{revtex4-1}

\usepackage{epstopdf}
\usepackage{color}
\usepackage{epsfig}
\usepackage{graphicx}
\usepackage{amsmath,amssymb,graphicx,amsthm,comment}

\renewcommand{\vec}[1]{\mathbf{#1}}

\newcommand{\pb}[1]{\left(#1\right)}

\newcommand{\bra}[1]{\left<#1\right|}                                           
\newcommand{\ket}[1]{\left|#1\right>}

\begin{document}


\title{Magnetoconductance signatures of chiral domain-wall bound states in
magnetic topological insulators}


\author{Kunal L. Tiwari}
\affiliation{Department of Physics, McGill University, Montr\'eal, Qu\'ebec,
Canada H3A 2T8}
\author{W.~A.~Coish}
\affiliation{Department of Physics, McGill University, Montr\'eal, Qu\'ebec,
Canada H3A 2T8}
\affiliation{Quantum Information Science Program, Canadian Institute for
Advanced Research, Toronto, Ontario M5G 1Z8, Canada}
\affiliation{Center for Quantum Devices and Station Q Copenhagen
Niels Bohr Institute, University of Copenhagen
2100 Copenhagen, Denmark}

\author{T. Pereg-Barnea}
\affiliation{Department of Physics, McGill University, Montr\'eal, Qu\'ebec,
Canada H3A 2T8}
\affiliation{Department of Condensed Matter Physics, Weizmann Institute of
    Science, Rehovot 7610001, Israel}

\date{\today}

\begin{abstract}
Recent magnetoconductance measurements performed on magnetic topological
insulator candidates have revealed butterfly-shaped hysteresis. 
This hysteresis has been attributed to the formation of gapless chiral
domain-wall bound states during a magnetic field sweep.  
We treat this phenomenon theoretically, providing a link between microscopic
magnetization dynamics and butterfly hysteresis in magnetoconductance.  
Further, we illustrate how a spatially resolved conductance measurement can
probe the most striking feature of the domain-wall bound states: their chirality. 
This work establishes a regime where a definitive link between butterfly
hysteresis in longitudinal magneto-conductance and domain-wall bound states can
be made.
This analysis provides an important tool for the identification
of magnetic topological insulators.
\end{abstract}






\maketitle


%

\section{Introduction}

\begin{figure}
 \includegraphics{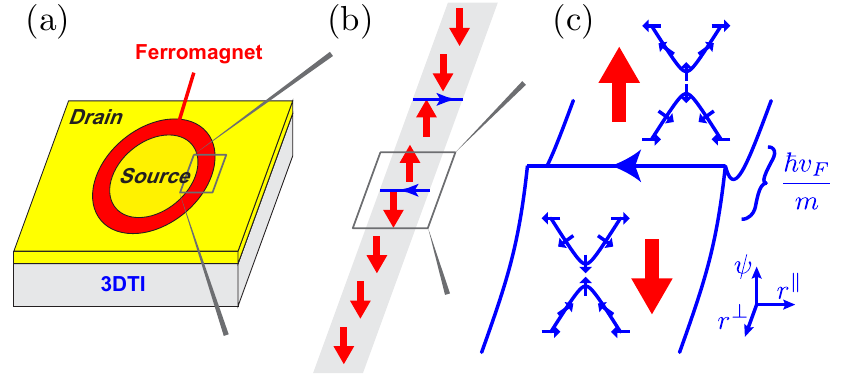}
 \caption{(a) Source and drain electrodes laid on the surface of a
 three-dimensional topological insulator having a ferromagnetic surface. (b)
 The magnetic subsystem is modeled as a one-dimensional Ising chain with
 periodic boundary conditions. Domain walls support chiral domain-wall bound
 states (DWBS's). (c) The DWBS's form at a domain wall, where the Dirac mass
 changes sign as the magnetization reverses.\label{introfig} }
 \end{figure}

The surface of a strong three-dimensional topological insulator is
characterized by an odd number of massless Dirac cones.\cite{FKM3DTI,MB3DTI}
The addition of static magnetic moments to the surface~\cite{Xu2012} leads to a local
Zeeman-like coupling term.  
If the moments are ferromagnetically coupled through a direct or indirect
[Ruderman–Kittel–Kasuya–Yosida (RKKY)] exchange,\cite{szRKKY} they become locked into finite domains of fixed
orientation at low temperature.
The Zeeman-like term then acts as a local Dirac mass with a sign determined by
the orientation of the proximal domain magnetization.  
Therefore, the electronic spectrum is gapped inside ferromagnetic domains. 
However, at domain boundaries, where the magnetization (and hence, the Dirac
mass) changes sign, the system hosts one-dimensional chiral edge states that
follow the domain walls.\cite{JackiwRebbi,JackiwMag,KMZEdgeStates,SSHSoliton}
These domain-wall bound states (DWBS's) are reminiscent of the chiral edge
modes in the quantum-Hall regime. 
The chirality of these conduction channels is determined by the magnetization
of the bounding domains and the spin-orbital structure of the surface states
[Fig.~\ref{introfig}(c), insets].
If the electronic chemical potential lies within the gap of the surface
spectrum, the low-temperature transport properties will be dominated by the
chiral and quantized conductance associated with DWBS's. 
In this way, electric conductance is determined by the magnetic configuration. 

DWBS's have been discussed in the experimental literature in the context of
hysteretic magnetoconductance in magnetic topological
insulators.\cite{Nakajima2016,OngJumps,CrSbTeParamag,Wang2016,checkelsky2012,
AQH_RIKEN, AQH_GHG}
In these experiments, a characteristic butterfly-shaped hysteresis is observed
as the applied field is increased, then decreased [see, e.g.,
Fig.~\ref{Egfig}(b), below, for an example arising from our model]. 
Butterfly hysteresis can be attributed to DWBS's by arguing that the
magnetization switches at the coercive field through the creation of domains,
resulting in a network of domain walls.
The DWBS's associated with the network of nucleated domains lead to excess
conductance.
Checkelsky et al.~\cite{checkelsky2012} have observed excess conductance as a
magnetic field is swept in Mn doped $\mathrm{Bi_2 (Te/Se)_3}$. 
Nakajima et al.~\cite{Nakajima2016} have observed similar magnetoconductance
hysteresis in $\mathrm{SmB_6}$, a candidate topological Kondo
insulator.\cite{DzeroTKI}
Wang et al.~\cite{Wang2016} have directly observed ferromagnetic domain
formation and growth using magnetic force microscopy in V doped
$\mathrm{Sb_2Te_3}$. The authors of that study observe \emph{suppressed}
(rather than enhanced) longitudinal conductance due to the dominant
contribution of bulk carriers to the magnetoconductance.
The picture of DWBS transport is compelling, but other mechanisms could explain
the observed magnetoconductance hysteresis. 
Analogous hysteresis curves are seen in distinctly non-topological systems, e.g.,
quantum-dot spin valves;\cite{NegTMR} similar tunneling magnetoresistance may
occur between domains of any conducting ferromagnet. 
Alternatively, paramagnetic cooling under a cycled magnetic
field~\cite{vlasov2017} in combination with a temperature-dependent conductance
could lead to similar hysteretic magnetotransport.

Previous theoretical studies of transport through DWBS's have focused on
current-driven magnetization dynamics.
These studies include spin-torque and associated domain-wall motion caused by
spin-polarized currents,\cite{AQHPiston, LossSpinTorque} and a proposal for
the inverse spin-galvanic effect as a mechanism for magnetization
switching.\cite{GarateCorbino}
In contrast, here we focus on the influence of magnetization dynamics on
magnetoconductance.
In particular, we directly link butterfly hysteresis in magnetoconductance to
DWBS dynamics driven by microscopic spin relaxation.  
Demonstrating that the hysteretic magnetoconductance is effected by chiral
transport channels would be definitive evidence of the DWBS hysteresis scenario.

This paper is organized as follows.
In Section~\ref{sec:model}, we introduce the model and a set of rate equations
governing magnetization dynamics under a magnetic field sweep.
In Section~\ref{sec:hysteresis}, we find an explicit closed-form analytical
expression for the trial-averaged magnetoconductance in a controlled limit and
predict its dependence on the magnetic-field sweep rate.
In Section~\ref{sec:spatcond}, we propose a spatially-resolved measurement capable of explicitly
probing the chirality of the DWBS's. 
In Section~\ref{sec:discussion}, we discuss the assumptions and limitations of
our proposal, concluding with a summary of our work. 

\section{model\label{sec:model}
}
We consider a topological surface state characterized by 
\begin{equation} H_e = v_\mathrm{F}\sum_{\vec k} c^\dagger_{\vec k}
    \left(k_x\sigma_y - k_y\sigma_x\right) c_{\vec{k}}.\label{eq:BareHam}
\end{equation}
Here, $\sigma_i$ are Pauli matrices and $c_{\vec k}=\left(c_{{\vec k}\uparrow},
c_{{\vec k}\downarrow}\right)^T$, where $c_{{\vec k}s}$ 
annihilates an electron with 2D surface momentum $\vec k$ and spin $s$.  
We assume the electronic system is in contact with a magnetic subsystem at the
surface.
The magnetic system induces a Zeeman-like coupling giving rise to a local Dirac
mass $m(\vec r)$ that is proportional to the local magnetization:
$\int\psi^\dagger(\vec r)m(\vec r)\sigma_z \psi(\vec r)\mathrm{d}\vec{r}$.  
Here, $\psi(\vec{r})=\sum_{\vec k} e^{i\vec{k}\cdot\vec{r}}c_{\vec
k}/\sqrt{A}$, and $A$ is the area of the 2D surface.
For a uniform magnetization (giving $m(\vec{r})=m$), this term gaps the surface
spectrum by $2m$.
When the magnetization switches (due, e.g., to a change in the ground state
during a magnetic-field sweep), it does so through the creation of local
domains, each associated with a bounding domain wall.
At a domain wall, where $m(\mathbf{r})$ changes sign, the electronic system
hosts a chiral DWBS.\cite{JackiwRebbi,JackiwMag}
Each chiral bound state extends a distance $\sim v_\mathrm{F}/m$ into the
magnetic domains [Fig.~\ref{introfig}(c)].  
Provided the domain walls are separated by more than this distance, there will
be no scattering between the bound states and each bound state will support a
single quantum of conductance $G_0=e^2/h$ in a direction determined by the
bounding magnetization [Fig.~\ref{introfig}(b)].
Electronic transport properties for such a system can be probed through
source and drain electrodes arranged in a Corbino geometry
[Fig.~\ref{introfig}(a)], providing evidence for the formation and dynamics of
chiral DWBS's.  
By considering a Corbino geometry, we avoid contributions from sample edges.
At sample edges, the projection of the magnetization onto the surface normal 
may change sign, leading to additional DWBS's~\cite{KMZEdgeStates} that complicate the
interpretation of magnetotransport measurements.

To model domain-wall dynamics and the resulting conductance during a
magnetic-field sweep for the geometry shown in Fig.~\ref{introfig}(a), we
consider a periodic Ising chain of $N$ spins with uniform coupling and a
uniform magnetic field, described by Hamiltonian:
\begin{align}
    H_m = -\sum_{i=1}^N \left(\frac{\xi}{2}\sigma_z^i \sigma_z^{i+1} + b
    \sigma_z^i \right)\label{eq:magham}.
\end{align}
Here, $\xi$ is the spin-spin coupling, $b$ is the Zeeman energy per spin due to
a magnetic field, and $\sigma_z^i$ is the Pauli operator acting on the spin
at site $i$.  The energy of the magnetic system is then fully characterized by
the magnetization $M$ (the difference between the number of spins up and
spins down) and by the number of domain walls $w$: $U = \xi w - b M$.

An infinite one-dimensional Ising system with short-range coupling has no
ferromagnetically ordered phase at finite temperature, due to the
logarithmically diverging entropic advantage in domain-wall
formation.\cite{Thouless1969} However, a \emph{finite} chain can order at
sufficiently low temperature.  For $b<0$, the ground state has $M=-N,\,w=0$,
while for $b>0$, the ground state has $M=N,\,w=0$.  As the magnetic field is
swept from negative to positive values at low temperature, the new ground state
can be reached through the production of transient chiral domain walls, $w\ne
0$.  For the geometry shown in Fig.~\ref{introfig}, each pair of domain walls
will be associated with one DWBS supporting a single quantum of conductance
from source to drain (and another conducting from drain to source),
Fig.~\ref{introfig}(b). This one-dimensional model directly addresses a
magnetic topological insulator in which individual magnetic impurities are
deposited around the Corbino annulus. To apply this model in the case of a
two-dimensional magnetic system, we require that the distance between the source
and drain electrodes be small compared to the typical domain size. At the same
time, this distance must be large enough that tunneling between the source and
drain is suppressed. The relevant distance is set by the decay length of
evanescent modes within a gapped region of the magnetic topological insulator
surface. This decay length is given by ${v_F}/{m}$.

In the limit of a slow sweep rate, $v = db/dt$ [cf.~Eq.~\eqref{eq:SlowSweep},
below], the magnetization reverses through the creation of only a single
counter-polarized domain associated with a single pair of domain walls.  
Magnetization dynamics can then be described in the subspace of $w=0,2$.  In
this subspace, $U$ is uniquely determined by $M$ ($M=\pm N$ correspond to
$w=0$, a single domain, while all other $M$ are reached for $w=2$).  We can
equivalently describe $U$ through the number of `up' spins, $n$: $M=2n-N$.  For
each $n$, the degenerate states of the magnetic system with energy $E_n$ are labeled
by a quantum number $\alpha$.

Transitions $n\alpha\to n'\alpha'$ in the magnetic subsystem (with rate
$\Gamma_{n\alpha\to n'\alpha'}$) result through the exchange of energy $\Delta
U_{n'n} = E_{n'}-E_n$ with an environment in thermal equilibrium and therefore obey detailed
balance: 
\begin{eqnarray}
    \Gamma_{n\alpha\to
    n'\alpha'}=\Gamma_{n'\alpha'\to n\alpha}e^{-\Delta U_{n'n}/k_\mathrm{B}T}.
\end{eqnarray}
For $T\to 0$, excitation processes $(\Delta U_{n'n}>0)$ are exponentially
suppressed, leaving relaxation only.
Restricting to single spin flips, the associated golden-rule
rates are then differentiated purely through the energy cost, $\Delta U_{n'n}$,
and the number of degenerate and accessible final states, $g_{n'}$: 
\begin{eqnarray}\Gamma_{n\to
    n'}\equiv\sum_{\alpha'} \Gamma_{n\alpha\to
    n'\alpha'} =g_{n'}\Gamma(\Delta U_{n'n}).  \label{eq:ratedefn}
\end{eqnarray}
In Appendix~\ref{app:rates}, we determine 
$g_{n'}$ and $\Gamma\pb{\Delta U_{n'n}}$ in terms of a microscopic system-environment
coupling.  The form of $\Gamma(\Delta U)$ depends generally on the coupling and
density-of-states of the environment.  For single spin flips, $\Delta M=\pm 2$
and the change in system energy is $\Delta U=\xi\Delta
w-b\Delta M$.  For $b>\xi$, there are three distinct relaxation rates:
domain-wall nucleation, $\Delta w = +2$, domain growth, $\Delta w=0$, and
domain-wall annihilation, $\Delta w=-2$. Thus, with $b>\xi$, the probability
to have $n$ spins up, $P_n$, obeys a simple classical (Pauli) master
equation:
\begin{eqnarray}
\dot{P}_n & = & \Gamma_{n-1\to n}P_{n-1}-\Gamma_{n\to n+1} P_{n};\,n=0,1,\ldots
    N,\label{eq:Pn}
\end{eqnarray}
where we set $\Gamma_{-1\to 0}=\Gamma_{N\to N+1}=0$ to apply Eq.~\eqref{eq:Pn}
for all $n$.
The distinct nonvanishing rates (for nucleation, growth, and annihilation,
respectively) are
\begin{align}
    &\Gamma_{0\to 1}=N\Gamma\textbf{(}2(\xi-b)\textbf{)}\equiv
    N\gamma_+ \\ 
    &\Gamma_{n\to n+1}=2 \Gamma(-2b)\equiv 2\gamma; 
    \hspace{10pt}n=1,\ldots,N-2\\ 
    &\Gamma_{N-1\to
    N}=\Gamma\textbf{(}-2(\xi+b)\textbf{)}\equiv \gamma_-.
\end{align}
The factor $g_1=N$ in the nucleation rate arises from the possibility to flip
any of the $N$ spins and the factor $g_{n'}=2$ (for $n'=2,\ldots,N-1$) is due to
the possibility to grow a domain by flipping a spin at either end.

\begin{figure}
\includegraphics{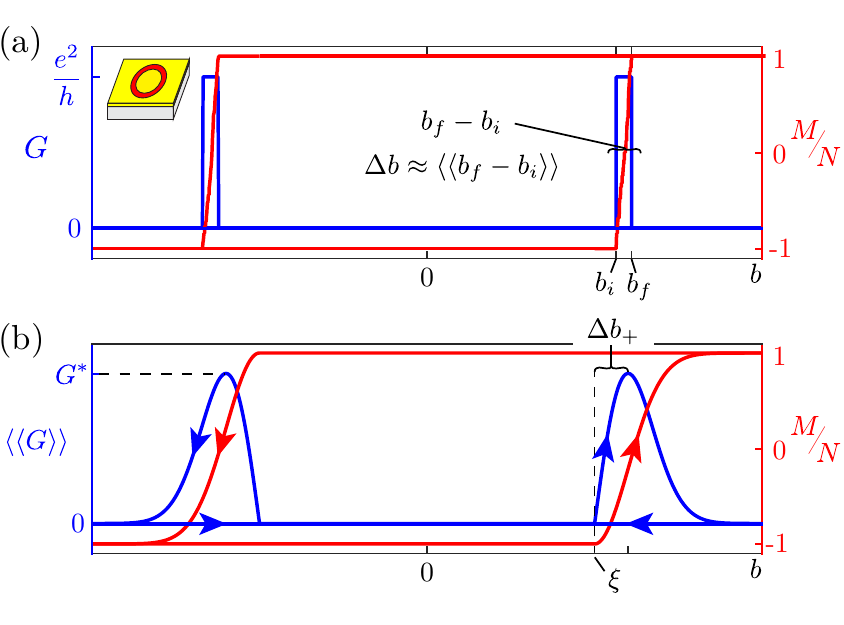}%
\caption{(a) Simulated conductance (blue) and magnetization (red) during a single field
sweep.  At $b=b_i > \xi$, a counter-polarized domain nucleates and grows
across the sample. At $b=b_f$, the domain has consumed the sample. While the
domain is growing, the conductance jumps to $G_0 = e^2/h$. The typical width of
a conductance plateau is $\Delta b$. (b) The trial-averaged conductance is
peaked at $b = \xi + \Delta b_+$ ($\Delta b_+\propto \sqrt{v}$), with maximum
value $G^*=G_0(\Delta b/\Delta b_+)\exp{\left(-1/2\right)}\propto\sqrt{v}$,
where $v=db/dt$ is the sweep rate.\label{Egfig}}
\end{figure}

During a field sweep at low temperature, the magnetic system will initially be in the
`down' state ($P_0=1$), and domain-wall nucleation may only occur above the 
coercive field ($b>\xi$).
For $b\gtrsim \xi$, we linearize the nucleation rate: 
\begin{align}\Gamma(\Delta U)\simeq
\Gamma^\prime |\Delta U|\Theta(-\Delta U)
\end{align} 
where $\Gamma' = - d\Gamma\pb{\Delta U = 0^-}/d\Delta U $.
This gives $\gamma_+\simeq
\Gamma'2(b-\xi)\Theta(b-\xi)$ (where $\Theta(\epsilon)$ is the Heaviside step
function).  
While $\gamma_+$ vanishes at $b=\xi$, the growth ($\gamma$) and annihilation
($\gamma_-$) rates remain finite and are taken to be approximately
$b$-independent over the range of interest.  These two rates are related within
the range of applicability of the linearization for $\Gamma(\Delta U)$:
$\gamma_-\simeq \Gamma(-4b)=2\gamma=2\Gamma(-2b)$.\footnote{By assuming this
relationship between $\gamma$ and $\gamma_-$, we simplify the solution to the
rate equation.  Moderate deviations in the annihilation rate $\gamma_-$ will
not significantly influence the predictions made here for the shape of
hysteresis curves resulting from $N\gg 1$ spin-flip events.}
We consider a linear sweep of $b$ with sweep rate $v$: $b=\xi+v t$, and
transform the equation-of-motion: $dP_n(t)/dt = v dP_n(b)/db$.  Integrating
Eq.~\eqref{eq:Pn} then gives (for $b>\xi$):
\begin{align}
    P_0(b) &= e^{-\frac{1}{2}\frac{(b-\xi)^2}{\Delta b_+^2}},\label{eq:P0}\\
P_N(b|b_i)&=\frac{\gamma\left(N-1,\frac{N(b-b_i)}{\Delta
b}\right)}{(N-2)!}\label{eq:Pnb}. 
\end{align} 
Here, $\gamma(s,y)=\int_0^y t^{s-1}e^{-t}dt$ is an incomplete gamma function,
$P_0(b)$ is the probability that the magnetic system remains in the initial state at
field strength $b$, and $P_N(b|b_i)$ is the conditional probability that all
$N$ spins have flipped provided that a nucleation event (initial spin flip)
occurred at field strength $b_i$.  We have introduced the scales 
\begin{align}\Delta b_+ = \sqrt{\frac{v}{2N\Gamma'}}\end{align} and 
\begin{align}\Delta b =
\frac{N v}{2\gamma}.\end{align}
These parameters have natural interpretations: $\Delta b_++\xi$ determines the typical field at
which the nucleation event occurs (Fig.~\ref{Egfig}(b)), and $\Delta b$ determines the
typical change in field strength during the growth of the domain
(Fig.~\ref{Egfig}(a)).

To guarantee that only a single pair of domain walls is created during
magnetization reversal (the single-pair limit), the typical time for domain
growth, $\sim\Delta b/v$, should be less than the typical time between 
nucleation events, $\sim\Delta b_+/v$: $\Delta b\ll\Delta b_+$, or equivalently:
\begin{equation}
v \ll \frac{2\gamma^2}{N^3 \Gamma'}.\label{eq:SlowSweep}
\end{equation}
If the linearization of the spin-relaxation rate is valid throughout $-2\xi \lesssim
\Delta U \leq 0 $, we can relate $\Gamma^{'}$ and $\gamma$ through
$\gamma = 2\Gamma^{'}\xi$  and the single-pair limit simplifies to $v \ll
\frac{2\xi\gamma}{N^3}$. We can substitute reasonable values for these
parameters to show that this condition may be realized experimentally. The
spin-relaxation rate, $\gamma$, is not well known for most magnetic topological
insulator candidates, but a range of values $\gamma = 1-10^{3}$ s$^{-1}$ is
routinely observed for single-electron spins in semiconductor quantum dots at
magnetic fields in the range of several Tesla.\cite{spinrelaxation} For
example, with $N=10$, $\gamma = 10^3$ s$^{-1}$, and a coercive field of 1 T
($\xi/\mu_Bg = 0.5$ T, where $g$ is the g-factor, and $\mu_B$ is the Bohr
magneton), the single-pair limit requires $v/\mu_Bg \ll 0.5$ T$\cdot$s$^{-1}$.

\section{magnetotransport hysteresis \label{sec:hysteresis}}

In the limit given by Eq.~\eqref{eq:SlowSweep}, we can use Eqs.~\eqref{eq:P0}
and~\eqref{eq:Pnb} to find closed-form
analytical expressions for the trial-averaged magnetoconductance.
There will be a single forward-conducting DWBS giving a single conductance
quantum ($G_0 =e^2/h$) for $0<n<N$, and none otherwise.
The trial-averaged (denoted $\langle\langle\cdots\rangle\rangle$) conductance
at field strength $b$ is then
\begin{align}
\langle \langle{G}\left( b \right)\rangle \rangle &= G_0
    \int_{b}^{\infty}db_f\int_{\xi}^{b} db_i  p_f\left(b_f|b_i\right)p_i
    \left(b_i\right).\label{eq:AverageConductance}
\end{align}
Here, $p_i\left(b_i\right) = -{\partial \over \partial b_i}P_0(b_i)$ is the
probability density that a nucleation event occurs at field strength $b_i$ and
$p_f(b_f|b_i)={\partial \over \partial b_f}{P}_N\left(b_f|b_i\right)$ is the
conditional probability density that an annihilation event occurs at $b_f$
given a nucleation event at $b_i$.
As the magnetic field is increased during a single sweep, the conductance
should show a jump to $G\simeq G_0$ starting at a typical field
$\left<\left<b_i\right>\right>=\xi+\Delta b_+$.
This jump coincides with the nucleation of a pair of domain walls following an
initial spin flip.
The conductance will then remain at its quantized value as the domain grows
through a sequence of spin flips at the domain walls. 
Finally, the conductance will return to zero when the final spin flips and the
magnetization has fully reversed. The final spin flip occurs at a typical field
$\left<\left<b_f\right>\right> \simeq \left<\left<b_i\right>\right>+\Delta b$
for $N\gg 1$ [Fig.~\ref{Egfig}(a)].  Although each sweep will be associated
with random values of $b_{i,f}$, averaging over many sweeps will result in a
robust averaged conductance with reproducible features, given by
Eq.~\eqref{eq:AverageConductance} [Fig.~\ref{Egfig}(b)].

The integrals in Eq.~\eqref{eq:AverageConductance} can be evaluated
approximately to leading order in the slow-field-sweep limit 
[Eq.~\eqref{eq:SlowSweep}, or equivalently $\Delta b/\Delta b_+\ll 1$],
giving a simple closed-form expression for the trial-averaged conductance:
\begin{align}
\langle \langle G\left(b\right) \rangle \rangle &\approx  G_0\frac{\Delta
b}{\Delta b_+}\frac{b-\xi}{\Delta b_+} e^{-\frac{1}{2}\left(\frac{b-\xi}{\Delta
b_+}\right)^{2}}\label{Gav}.
\end{align}
The averaged conductance is peaked at $b = \xi + \Delta b_+$, with maximal
value $G^*= G_0(\Delta b/\Delta b_+)\exp \left(-1/2\right)$.
The trial-averaged conductance [from Eq.~\eqref{Gav}] and magnetization are
plotted in Fig.~\ref{Egfig}(b).  
A measurement of the correlated change in conductance and magnetization would
provide strong evidence for the DWBS picture studied here.  
Even in the absence of microscopic time-resolved magnetization measurements,
the connection between transport and magnetization dynamics can be verified
from Eq.~\eqref{Gav} through transport alone.
In particular, Eq.~\eqref{Gav} predicts nontrivial dependences for the averaged
conductance peak height ($G^*\propto \sqrt{v}$) and the maximum position
($\Delta b_+\propto \sqrt{v}$) as the sweep rate $v$ is varied.  
The specific dependences predicted here are non-universal, relying on a
linearization of the rates, $\Gamma(\Delta U)\propto |\Delta U|$.  
More generally, for a typical power-law form $\Gamma(\Delta U)\propto |\Delta
U|^\eta$, we find a modified lineshape giving a conductance maximum $G^*\propto
v^{\eta/(\eta+1)}$ and peak width $\Delta b_+\propto v^{1/(\eta+1)}$ (see
Appendix~\ref{app:etagen} for the detailed forms of $\Delta b_+$ and $G^*$ in this
general case).
Experimental confirmation of these dependencies would be strong evidence for
the magnetic origin of the hysteretic conductance and could help to establish
the relevant spin-relaxation mechanisms through the exponent $\eta$.

\section{Spatially resolved conductance and bound-state chirality\label{sec:spatcond}}

\begin{figure}\definecolor{mygreen}{RGB}{0,200,0}
\includegraphics{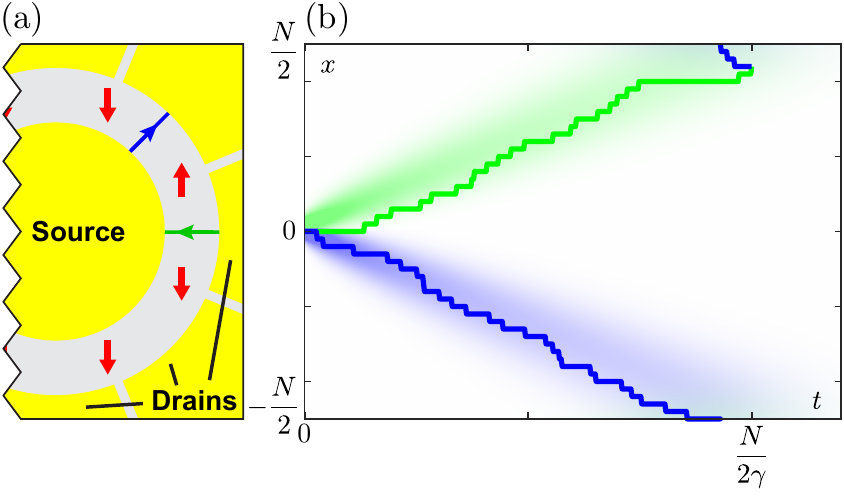}%
\caption{\label{spatcondfig} (a) Segmented gates can be used in the Corbino
geometry to spatially resolve bias-dependent conductance from domain-wall
bound states (DWBS's) to establish their chirality. (b) Probability density
for the outward-conducting (blue) and inward-conducting (green) DWBS's
nucleating at $x=0$ and diffusing via a directed random walk before
annihilating at $|x|\simeq N/2$ for $N\gg 1$ spins.  To produce this plot, we
have taken $N=50\gg 1$ spins, leading to well-resolved trajectories.}
\end{figure}

The calculation presented in the previous section could be used to connect
microscopic magnetization dynamics to hysteretic conductance.  However, it does
not directly address the chirality of the associated transport channels.  To
establish chirality, it may be useful to consider a configuration where the
DWBS's can be spatially resolved by probing conductance as a function of the
bias direction and position around the Corbino annulus with segmented
electrodes [Fig.~\ref{spatcondfig}(a)].  Such a measurement would reveal a
conductance peak at one side of a domain for outward bias, and at the opposite
side for inward bias.  When only a single pair of domain walls is produced, it
should be possible to resolve the two domain-wall-induced conductance
peaks---they will be separated by a maximum distance comparable to the sample
size.

To flesh out the domain-wall dynamics, we now consider an experiment where the
ground state is prepared with all spins down at $b<0$, then the magnetic field
is rapidly pulsed to a fixed value $b>\xi$, giving a nonzero nucleation rate
and a nonzero spin-flip rate at either end of a nucleated domain, 
$\gamma=\Gamma(-2b)$.  
As before, we can describe the dynamics of magnetization reversal in the
single-pair limit.
After a nucleation event, the position of a domain wall is described, through
Eq.~\eqref{eq:Pn}, by a directed random walk with Poissonian-distributed step
times at rate $\gamma$.  
The probability that there is a domain wall $x$ sites away from the nucleation
site a time $t$ after the nucleation event, $P(x,t)$, is then well-approximated
by a gamma distribution for $t\lesssim N/2\gamma$ (before an annihilation event):
\begin{equation}
P\left(x,t\right) \simeq (\gamma t)^{|x|-1}e^{-\gamma
t}/(|x|-1)! \label{spatcondprob}.
\end{equation}
In this experiment, one would record the inward (for reverse source-drain bias)
and outward (for forward source-drain bias) conductances as a
function of position and time. The conductance for a given bias direction will be peaked
at the position of the relevant domain wall. As a function of time, this
conductance peak will move around the Corbino annulus in a direction dictated
by the chirality of the underlying DWBS\@. In Fig.~\ref{spatcondfig}(b), we show
the probability $P(x,t)$ for the inward-conducting (green shading) and
outward-conducting (blue shading) DWBS's. We have also plotted simulated
specific trajectories for a single run of the experiment (solid lines). These
correspond to the position of the conductance peaks for inward and outward
bias. The connection between dynamics of the conductance and underlying
magnetization could potentially be verified from Eq.~\eqref{spatcondprob} by
modifying the end value of the magnetic field, thus varying
$\gamma=\Gamma(-2b)$ and the associated distribution of trajectories. 

\section{Discussion and conclusions\label{sec:discussion}}
Several important assumptions and approximations have been
made throughout this work.  
The focus here on the Corbino geometry avoids complications associated with
spurious edge transport for other geometries, but most of our analysis would be 
valid for a finite linear geometry with source and drain leads placed close
together.
Here, the domain-wall annihilation is replaced by an `escape' of domain
walls from the region between the leads.  
Another crucial simplification of our analysis is to focus on the low-temperature
limit, where excitation processes are exponentially suppressed.  
This has allowed us to provide simple closed-form expressions for the
trial-averaged conductance and probability distribution for domain-wall
diffusion under magnetization dynamics.  
It would also be interesting to study the link between magnetization dynamics
and transport for a quantum magnet (beyond the Ising limit taken here)---there,
the delocalized eigenstates (e.g.\ spin waves) could show an interesting
interplay with the localized conductance measurement illustrated in
Fig.~\ref{spatcondfig}(a).
Such a study is, however, beyond the scope of this work.

The analysis presented here may provide an important tool to rule out possible
alternative explanations for hysteretic magnetotransport (including
paramagnetic cooling from field cycling~\cite{vlasov2017} and
temperature-dependent resistance, or effects due to tunneling magnetoresistance).  
The specific form of the trial-averaged conductance given in
Eq.~\eqref{eq:AverageConductance} and, in particular, its dependence on the
sweep rate $v=db/dt$ predicted here would provide a strong link between
magnetization dynamics and conductance via DWBS's.  

In summary, we are able to provide simple analytical formulas for the expected
butterfly-shaped hysteretic conductance, including the dependence of key
features on the sweep rate in a controlled limit.  
These features can be used to directly confirm the presence and dynamics of
DWBS's and their chirality, addressing an important question in the
experimental identification of magnetic topological insulators.

\begin{acknowledgments}
We thank Guillaume Gervais and Johnpierre Paglione for useful discussions.  
We acknowledge support from NSERC, FRQNT, INTRIQ, CIFAR, Nordea Fonden, and the
Schulich Graduate Fellowship. 
\end{acknowledgments}

\appendix
\section{Microscopic spin-flip rate\label{app:rates}}
In this appendix, we show how the spin-flip rates used in the main text arise
from a coupling between the magnetic system and an environment. 

We consider a Hamiltonian for the magnetic system and environment of the form
\begin{align}
    H = H_m + H_B + V.
\end{align}
Here, $H_m$ is the magnetic system Hamiltonian given in Eq.~\eqref{eq:magham},
$H_B$ is the environment Hamiltonian and $V$ is the coupling between system and
environment.  As in the main text, we work in the subspace spanned by eigenstates of
the magnetic system Hamiltonian with only zero or two domain walls. We can denote such
eigenstates
\begin{align}
    H_m\ket{n,\alpha} = E_n \ket{n,\alpha},
\end{align}
where $n$ is again the number of upward-oriented spins in the system and
$\alpha$ labels the degenerate states with energy $E_n$.
The subspaces with $n = 0,N$ each contain only a single spin configuration,
while those with $n
= 1,\dots,N-1$ each host $N$ degenerate configurations. 
We consider a coupling of the form
\begin{align}
    V = \sum_{j}\sigma_-^{j}B_j + H.c., 
\end{align}
where $\sigma_{\pm}^j$ flips the spin at site $j$, and $B_j$ is an operator
that acts in the environment subspace.
\newcommand{\nap}{n' \alpha'}
\newcommand{\na}{n\alpha}

Our goal is to determine the transition rate, $\Gamma_{n\rightarrow n'}$,
from a particular state with $n$ spins up to any of the $g_{n'}$ states with
$n'$ spins up.  
This may be found from $P_{\na\rightarrow\nap}$, the probability for a
transition from $\ket{\na}$ to $\ket{\nap}$ calculated to second order in
time-dependent perturbation theory: 
\newcommand{\Unpn}{\Delta U_{n'n}}
\begin{align}
    \Gamma_{n\rightarrow n'} =\sum_{\alpha'}\lim_{t\rightarrow\infty} 
    \frac{d}{dt} P_{n\alpha \rightarrow n' \alpha'}\pb{t}. \label{eq:Gammasetup}
\end{align}
As described in the main text, we take an environment initially in thermal
equilibrium at temperature $T$.  In the low-$T$ limit, excitation is
exponentially suppressed leaving relaxation rates only (having $\Delta
U_{n'n}=E_{n'}-E_n<0$).  We focus on transitions above the coercive field
during a magnetic field up-sweep, where relaxation occurs only through
spin-flips from down to up. Direct evaluation of Eq.~\eqref{eq:Gammasetup} in
this regime gives the relaxation rates:
\begin{equation}
\Gamma_{n\rightarrow n'}=g_{n'}\Gamma(\Delta U_{n'n}),
\end{equation}
where $g_{n'} = \sum_{j\alpha'} |\bra{n'\alpha'}\sigma_+^{j}\ket{n\alpha}|^2$
gives the number of accessible final states, and where the single-spin
relaxation rate is
\begin{equation}\label{eq:GammaDeltaU}
\Gamma(\Delta U) = \int_{-\infty}^\infty dt e^{-i\Delta U t -0^+|t|}\langle
B^\dagger(t) B\rangle_0
\end{equation} 
Here, $B^\dagger(t)=e^{i H_B t}B^\dagger e^{-i H_B t}$ describes the
interaction picture, $\langle \cdots\rangle_0=\bra{0}\cdots\ket{0}$ is an
average with respect to the ground state of the environment Hamiltonian $H_B$,
and $0^+$ is a positive infinitesimal.  In writing Eq.~\eqref{eq:GammaDeltaU},
we have assumed that the environment acting at each site is uncorrelated and
equivalent at all sites, i.e.: $\langle
B_i^\dagger(t)B_j\rangle=\delta_{ij}\langle B^\dagger(t)B\rangle$.

Expanding Eq.~\eqref{eq:GammaDeltaU} in terms of environment eigenstates
($H_B\ket{k}=\epsilon_k\ket{k}$; $\epsilon_k\ge 0$) gives the standard Fermi's
golden rule result:
\begin{equation}\label{eq:GoldenRule}
\Gamma(\Delta U) = 2\pi \sum_k |B_{0k}|^2\delta(\Delta U+\epsilon_k),
\end{equation}  
where $B_{0k}=\bra{0}B\ket{k}$.  In many cases, the environment coupling can be
described (at least approximately) as a unique function of the energy,
$B_{0k}=g(\epsilon_k)$.  Taking the continuum limit of
Eq.~\eqref{eq:GoldenRule} then gives:
\begin{equation}\label{eq:GoldenRuleContinuum}
\Gamma(\Delta U)=2\pi D(-\Delta U)|g(-\Delta U)|^2,
\end{equation}  
with environment density of excitations $D(\epsilon)=\sum_k
\delta(\epsilon-\epsilon_k)$.

In general, the energy-dependence of Eq.~\eqref{eq:GoldenRuleContinuum} will be
determined by both the density of excitations in the environment,
$D(\epsilon)$, and the coupling $g(\epsilon)$.  If the low-energy excitations
are long-wavelength phonons or magnons, the density of states will vanish at
low energy, typically like $D(\epsilon)\propto \epsilon^{d-1}$ in $d=2,3$
dimensions.  Provided the coupling $g(\epsilon)$ does not diverge as
$\epsilon\to 0$, this will result in a rate that vanishes typically with some
power-law: $\Gamma(\Delta U)\propto |\Delta U|^\eta$.  We consider a general
power-law in Appendix~\ref{app:etagen}, below.  In the main text, we considered
a rate that is linearizable at low energy.  Such a linearizable rate arises
naturally, e.g., when spin-flips occur due to cotunneling with a metallic
reservoir at low temperature.  These second-order tunneling processes give rise
to an approximately energy-independent effective coupling $g(\epsilon)\sim
t_c^2/U_c$ with tunnel coupling $t_c$ and effective charging energy $U_c$.  The
environmental excitations are electron-hole pairs created about the Fermi level
in an energy window of size $|\Delta U|$, leading to an overall linear energy
dependence at low temperature: $\Gamma(\Delta U)\propto |\Delta U|$ [see, e.g.,
Eq.~(6) in Ref.~\onlinecite{qassemi2009}]. The spin-relaxation mechanisms in
magnetic topological insulator candidates are not well known. Our proposed
measurement should help identify the relevant mechanisms through determination
of the parameter $\eta$.

\section{Magnetization dynamics for a generalized spin-flip rate\label{app:etagen}}
In the main text, we linearlized the nucleation rate, $\Gamma ( \Delta U)
\simeq \Gamma ' |\Delta U|$, above the coercive field.
In general, the nucleation rate could have an arbitrary
power-law dependence close to the coercive field, $\Gamma (\Delta U) \propto
|\Delta U|^\eta$, with exponent $\eta$ depending on the reservoir
spectral density:
\begin{align}
    \gamma_+ \simeq \Gamma^{(\eta)}\textbf{(}2(b-\xi)\textbf{)}^\eta\Theta(b-\xi)
\end{align}
The growth and annihilation rates are again taken to be constant about the
coercive field.
Using this general $\gamma_+$ we find 
\begin{align}
    P_0(b) = \exp\left[-\frac{\eta}{\eta+1}\left(\frac{b-\xi}{\Delta
    b_+}\right)^{\eta+1}\right]\label{eq:appP0}.
\end{align}
The conditional probability that all $N$ spins have flipped after a nucleation
event, $P_N(b|b_i)$, is unchanged relative to Eq.~\eqref{eq:Pnb} of the main text
since it depends only on $\gamma$ and
$\gamma_-$. We have introduced the modified scale 
\begin{align}
    \Delta b_+ = \left(\frac{v\eta}{2^\eta N
    \Gamma^{(\eta)}}\right)^{\frac{1}{\eta +1 }}.
\end{align}

The condition to guarantee that the magnetization reversal occurs through only a
single nucleation event is again that the typical time for domain growth, $\sim
\Delta b/v$ should be small compared to the typical time between nucleation
events (now $\sim \Delta b_+ / \eta v$). 
The more general single-pair limit is then
\begin{align}
    v \ll \eta \left(\frac{2     \gamma^{\eta+1}
    }{N^{\eta+2}\Gamma^{(\eta)}}\right)^{\frac{1}{\eta}}.
\end{align}

We evaluate the trial-averaged conductance integral,
Eq.~\eqref{eq:AverageConductance}, using Eq.~\eqref{eq:appP0} in the
slow-sweep-limit to find
\begin{align}
    \langle\langle G(b)\rangle\rangle \simeq G_0 \eta \frac{\Delta b}{\Delta
    b_+}\left(\frac{b-\xi}{\Delta b_+}\right)^{\eta} \exp\left[-\frac{\eta}{\eta+1}\left(\frac{b-\xi}{\Delta
    b_+}\right)^{\eta+1}\right].
\end{align}
The average conductance is again peaked at $b=\xi + \Delta b_+$, with maximal
value \begin{align}G^* = G_0\eta\frac{\Delta b}{\Delta b_+}\exp
\left[-\frac{\eta}{\eta+1}\right].\end{align}
The $v$ scaling presented in the main text is found by examining the
$v$-dependence of $\Delta b_+$ and $G^*$. $\Delta b_+$ is proportional to
$v^{\frac{1}{\eta +1}}$. The $v$-dependence in $G^*$ comes from the ratio
$\Delta b / \Delta b_+$. $\Delta b$ is proportional to $v$. Substituting the
$v$-dependence of $\Delta b_+$, we find $G^*\propto v^{\frac{\eta}{\eta+1}}$.

\bibliography{AQHBib_WAC}

\begin{thebibliography}{25}%
\makeatletter
\providecommand \@ifxundefined [1]{%
 \@ifx{#1\undefined}
}%
\providecommand \@ifnum [1]{%
 \ifnum #1\expandafter \@firstoftwo
 \else \expandafter \@secondoftwo
 \fi
}%
\providecommand \@ifx [1]{%
 \ifx #1\expandafter \@firstoftwo
 \else \expandafter \@secondoftwo
 \fi
}%
\providecommand \natexlab [1]{#1}%
\providecommand \enquote  [1]{``#1''}%
\providecommand \bibnamefont  [1]{#1}%
\providecommand \bibfnamefont [1]{#1}%
\providecommand \citenamefont [1]{#1}%
\providecommand \href@noop [0]{\@secondoftwo}%
\providecommand \href [0]{\begingroup \@sanitize@url \@href}%
\providecommand \@href[1]{\@@startlink{#1}\@@href}%
\providecommand \@@href[1]{\endgroup#1\@@endlink}%
\providecommand \@sanitize@url [0]{\catcode `\\12\catcode `\$12\catcode
  `\&12\catcode `\#12\catcode `\^12\catcode `\_12\catcode `\%12\relax}%
\providecommand \@@startlink[1]{}%
\providecommand \@@endlink[0]{}%
\providecommand \url  [0]{\begingroup\@sanitize@url \@url }%
\providecommand \@url [1]{\endgroup\@href {#1}{\urlprefix }}%
\providecommand \urlprefix  [0]{URL }%
\providecommand \Eprint [0]{\href }%
\providecommand \doibase [0]{http://dx.doi.org/}%
\providecommand \selectlanguage [0]{\@gobble}%
\providecommand \bibinfo  [0]{\@secondoftwo}%
\providecommand \bibfield  [0]{\@secondoftwo}%
\providecommand \translation [1]{[#1]}%
\providecommand \BibitemOpen [0]{}%
\providecommand \bibitemStop [0]{}%
\providecommand \bibitemNoStop [0]{.\EOS\space}%
\providecommand \EOS [0]{\spacefactor3000\relax}%
\providecommand \BibitemShut  [1]{\csname bibitem#1\endcsname}%
\let\auto@bib@innerbib\@empty
\bibitem [{\citenamefont {Fu}\ \emph {et~al.}(2007)\citenamefont {Fu},
  \citenamefont {Kane},\ and\ \citenamefont {Mele}}]{FKM3DTI}%
  \BibitemOpen
  \bibfield  {author} {\bibinfo {author} {\bibfnamefont {L.}~\bibnamefont
  {Fu}}, \bibinfo {author} {\bibfnamefont {C.~L.}\ \bibnamefont {Kane}}, \ and\
  \bibinfo {author} {\bibfnamefont {E.~J.}\ \bibnamefont {Mele}},\ }\href
  {\doibase 10.1103/PhysRevLett.98.106803} {\bibfield  {journal} {\bibinfo
  {journal} {Phys. Rev. Lett.}\ }\textbf {\bibinfo {volume} {98}},\ \bibinfo
  {pages} {106803} (\bibinfo {year} {2007})}\BibitemShut {NoStop}%
\bibitem [{\citenamefont {Moore}\ and\ \citenamefont {Balents}(2007)}]{MB3DTI}%
  \BibitemOpen
  \bibfield  {author} {\bibinfo {author} {\bibfnamefont {J.~E.}\ \bibnamefont
  {Moore}}\ and\ \bibinfo {author} {\bibfnamefont {L.}~\bibnamefont
  {Balents}},\ }\href {\doibase 10.1103/PhysRevB.75.121306} {\bibfield
  {journal} {\bibinfo  {journal} {Phys. Rev. B}\ }\textbf {\bibinfo {volume}
  {75}},\ \bibinfo {pages} {121306} (\bibinfo {year} {2007})}\BibitemShut
  {NoStop}%
\bibitem [{\citenamefont {Xu}\ \emph {et~al.}(2012)\citenamefont {Xu},
  \citenamefont {Neupane}, \citenamefont {Liu}, \citenamefont {Zhang},
  \citenamefont {Richardella}, \citenamefont {Andrew~Wray}, \citenamefont
  {Alidoust}, \citenamefont {Leandersson}, \citenamefont {Balasubramanian},
  \citenamefont {Sanchez-Barriga}, \citenamefont {Rader}, \citenamefont
  {Landolt}, \citenamefont {Slomski}, \citenamefont {Hugo~Dil}, \citenamefont
  {Osterwalder}, \citenamefont {Chang}, \citenamefont {Jeng}, \citenamefont
  {Lin}, \citenamefont {Bansil}, \citenamefont {Samarth},\ and\ \citenamefont
  {Zahid~Hasan}}]{Xu2012}%
  \BibitemOpen
  \bibfield  {author} {\bibinfo {author} {\bibfnamefont {S.-Y.}\ \bibnamefont
  {Xu}}, \bibinfo {author} {\bibfnamefont {M.}~\bibnamefont {Neupane}},
  \bibinfo {author} {\bibfnamefont {C.}~\bibnamefont {Liu}}, \bibinfo {author}
  {\bibfnamefont {D.}~\bibnamefont {Zhang}}, \bibinfo {author} {\bibfnamefont
  {A.}~\bibnamefont {Richardella}}, \bibinfo {author} {\bibfnamefont
  {L.}~\bibnamefont {Andrew~Wray}}, \bibinfo {author} {\bibfnamefont
  {N.}~\bibnamefont {Alidoust}}, \bibinfo {author} {\bibfnamefont
  {M.}~\bibnamefont {Leandersson}}, \bibinfo {author} {\bibfnamefont
  {T.}~\bibnamefont {Balasubramanian}}, \bibinfo {author} {\bibfnamefont
  {J.}~\bibnamefont {Sanchez-Barriga}}, \bibinfo {author} {\bibfnamefont
  {O.}~\bibnamefont {Rader}}, \bibinfo {author} {\bibfnamefont
  {G.}~\bibnamefont {Landolt}}, \bibinfo {author} {\bibfnamefont
  {B.}~\bibnamefont {Slomski}}, \bibinfo {author} {\bibfnamefont
  {J.}~\bibnamefont {Hugo~Dil}}, \bibinfo {author} {\bibfnamefont
  {J.}~\bibnamefont {Osterwalder}}, \bibinfo {author} {\bibfnamefont {T.-R.}\
  \bibnamefont {Chang}}, \bibinfo {author} {\bibfnamefont {H.-T.}\ \bibnamefont
  {Jeng}}, \bibinfo {author} {\bibfnamefont {H.}~\bibnamefont {Lin}}, \bibinfo
  {author} {\bibfnamefont {A.}~\bibnamefont {Bansil}}, \bibinfo {author}
  {\bibfnamefont {N.}~\bibnamefont {Samarth}}, \ and\ \bibinfo {author}
  {\bibfnamefont {M.}~\bibnamefont {Zahid~Hasan}},\ }\href {\doibase
  10.1038/nphys2351} {\bibfield  {journal} {\bibinfo  {journal} {Nat Phys}\
  }\textbf {\bibinfo {volume} {8}},\ \bibinfo {pages} {616} (\bibinfo {year}
  {2012})}\BibitemShut {NoStop}%
\bibitem [{\citenamefont {Liu}\ \emph {et~al.}(2009)\citenamefont {Liu},
  \citenamefont {Liu}, \citenamefont {Xu}, \citenamefont {Qi},\ and\
  \citenamefont {Zhang}}]{szRKKY}%
  \BibitemOpen
  \bibfield  {author} {\bibinfo {author} {\bibfnamefont {Q.}~\bibnamefont
  {Liu}}, \bibinfo {author} {\bibfnamefont {C.-X.}\ \bibnamefont {Liu}},
  \bibinfo {author} {\bibfnamefont {C.}~\bibnamefont {Xu}}, \bibinfo {author}
  {\bibfnamefont {X.-L.}\ \bibnamefont {Qi}}, \ and\ \bibinfo {author}
  {\bibfnamefont {S.-C.}\ \bibnamefont {Zhang}},\ }\href {\doibase
  10.1103/PhysRevLett.102.156603} {\bibfield  {journal} {\bibinfo  {journal}
  {Phys. Rev. Lett.}\ }\textbf {\bibinfo {volume} {102}},\ \bibinfo {pages}
  {156603} (\bibinfo {year} {2009})}\BibitemShut {NoStop}%
\bibitem [{\citenamefont {Jackiw}\ and\ \citenamefont
  {Rebbi}(1976)}]{JackiwRebbi}%
  \BibitemOpen
  \bibfield  {author} {\bibinfo {author} {\bibfnamefont {R.}~\bibnamefont
  {Jackiw}}\ and\ \bibinfo {author} {\bibfnamefont {C.}~\bibnamefont {Rebbi}},\
  }\href {\doibase 10.1103/PhysRevD.13.3398} {\bibfield  {journal} {\bibinfo
  {journal} {Phys. Rev. D}\ }\textbf {\bibinfo {volume} {13}},\ \bibinfo
  {pages} {3398} (\bibinfo {year} {1976})}\BibitemShut {NoStop}%
\bibitem [{\citenamefont {Jackiw}(1984)}]{JackiwMag}%
  \BibitemOpen
  \bibfield  {author} {\bibinfo {author} {\bibfnamefont {R.}~\bibnamefont
  {Jackiw}},\ }\href {\doibase 10.1103/PhysRevD.29.2375} {\bibfield  {journal}
  {\bibinfo  {journal} {Phys. Rev. D}\ }\textbf {\bibinfo {volume} {29}},\
  \bibinfo {pages} {2375} (\bibinfo {year} {1984})}\BibitemShut {NoStop}%
\bibitem [{\citenamefont {Zhang}\ \emph {et~al.}(2013)\citenamefont {Zhang},
  \citenamefont {Kane},\ and\ \citenamefont {Mele}}]{KMZEdgeStates}%
  \BibitemOpen
  \bibfield  {author} {\bibinfo {author} {\bibfnamefont {F.}~\bibnamefont
  {Zhang}}, \bibinfo {author} {\bibfnamefont {C.~L.}\ \bibnamefont {Kane}}, \
  and\ \bibinfo {author} {\bibfnamefont {E.~J.}\ \bibnamefont {Mele}},\ }\href
  {\doibase 10.1103/PhysRevLett.110.046404} {\bibfield  {journal} {\bibinfo
  {journal} {Phys. Rev. Lett.}\ }\textbf {\bibinfo {volume} {110}},\ \bibinfo
  {pages} {046404} (\bibinfo {year} {2013})}\BibitemShut {NoStop}%
\bibitem [{\citenamefont {Su}\ \emph {et~al.}(1980)\citenamefont {Su},
  \citenamefont {Schrieffer},\ and\ \citenamefont {Heeger}}]{SSHSoliton}%
  \BibitemOpen
  \bibfield  {author} {\bibinfo {author} {\bibfnamefont {W.~P.}\ \bibnamefont
  {Su}}, \bibinfo {author} {\bibfnamefont {J.~R.}\ \bibnamefont {Schrieffer}},
  \ and\ \bibinfo {author} {\bibfnamefont {A.~J.}\ \bibnamefont {Heeger}},\
  }\href {\doibase 10.1103/PhysRevB.22.2099} {\bibfield  {journal} {\bibinfo
  {journal} {Phys. Rev. B}\ }\textbf {\bibinfo {volume} {22}},\ \bibinfo
  {pages} {2099} (\bibinfo {year} {1980})}\BibitemShut {NoStop}%
\bibitem [{\citenamefont {Nakajima}\ \emph {et~al.}(2016)\citenamefont
  {Nakajima}, \citenamefont {Syers}, \citenamefont {Wang}, \citenamefont
  {Wang},\ and\ \citenamefont {Paglione}}]{Nakajima2016}%
  \BibitemOpen
  \bibfield  {author} {\bibinfo {author} {\bibfnamefont {Y.}~\bibnamefont
  {Nakajima}}, \bibinfo {author} {\bibfnamefont {P.}~\bibnamefont {Syers}},
  \bibinfo {author} {\bibfnamefont {X.}~\bibnamefont {Wang}}, \bibinfo {author}
  {\bibfnamefont {R.}~\bibnamefont {Wang}}, \ and\ \bibinfo {author}
  {\bibfnamefont {J.}~\bibnamefont {Paglione}},\ }\href
  {http://dx.doi.org/10.1038/nphys3555} {\bibfield  {journal} {\bibinfo
  {journal} {Nat Phys}\ }\textbf {\bibinfo {volume} {12}},\ \bibinfo {pages}
  {213} (\bibinfo {year} {2016})}\BibitemShut {NoStop}%
\bibitem [{\citenamefont {Liu}\ \emph {et~al.}(2016)\citenamefont {Liu},
  \citenamefont {Wang}, \citenamefont {Richardella}, \citenamefont {Kandala},
  \citenamefont {Li}, \citenamefont {Yazdani}, \citenamefont {Samarth},\ and\
  \citenamefont {Ong}}]{OngJumps}%
  \BibitemOpen
  \bibfield  {author} {\bibinfo {author} {\bibfnamefont {M.}~\bibnamefont
  {Liu}}, \bibinfo {author} {\bibfnamefont {W.}~\bibnamefont {Wang}}, \bibinfo
  {author} {\bibfnamefont {A.~R.}\ \bibnamefont {Richardella}}, \bibinfo
  {author} {\bibfnamefont {A.}~\bibnamefont {Kandala}}, \bibinfo {author}
  {\bibfnamefont {J.}~\bibnamefont {Li}}, \bibinfo {author} {\bibfnamefont
  {A.}~\bibnamefont {Yazdani}}, \bibinfo {author} {\bibfnamefont
  {N.}~\bibnamefont {Samarth}}, \ and\ \bibinfo {author} {\bibfnamefont
  {N.~P.}\ \bibnamefont {Ong}},\ }\href {\doibase 10.1126/sciadv.1600167} {\
  \textbf {\bibinfo {volume} {2}} (\bibinfo {year} {2016}),\
  10.1126/sciadv.1600167}\BibitemShut {NoStop}%
\bibitem [{\citenamefont {Lachman}\ \emph {et~al.}(2015)\citenamefont
  {Lachman}, \citenamefont {Young}, \citenamefont {Richardella}, \citenamefont
  {Cuppens}, \citenamefont {Naren}, \citenamefont {Anahory}, \citenamefont
  {Meltzer}, \citenamefont {Kandala}, \citenamefont {Kempinger}, \citenamefont
  {Myasoedov} \emph {et~al.}}]{CrSbTeParamag}%
  \BibitemOpen
  \bibfield  {author} {\bibinfo {author} {\bibfnamefont {E.~O.}\ \bibnamefont
  {Lachman}}, \bibinfo {author} {\bibfnamefont {A.~F.}\ \bibnamefont {Young}},
  \bibinfo {author} {\bibfnamefont {A.}~\bibnamefont {Richardella}}, \bibinfo
  {author} {\bibfnamefont {J.}~\bibnamefont {Cuppens}}, \bibinfo {author}
  {\bibfnamefont {H.}~\bibnamefont {Naren}}, \bibinfo {author} {\bibfnamefont
  {Y.}~\bibnamefont {Anahory}}, \bibinfo {author} {\bibfnamefont {A.~Y.}\
  \bibnamefont {Meltzer}}, \bibinfo {author} {\bibfnamefont {A.}~\bibnamefont
  {Kandala}}, \bibinfo {author} {\bibfnamefont {S.}~\bibnamefont {Kempinger}},
  \bibinfo {author} {\bibfnamefont {Y.}~\bibnamefont {Myasoedov}},  \emph
  {et~al.},\ }\href@noop {} {\bibfield  {journal} {\bibinfo  {journal} {Science
  advances}\ }\textbf {\bibinfo {volume} {1}},\ \bibinfo {pages} {e1500740}
  (\bibinfo {year} {2015})}\BibitemShut {NoStop}%
\bibitem [{\citenamefont {Wang}\ \emph {et~al.}(2016)\citenamefont {Wang},
  \citenamefont {Chang}, \citenamefont {Moodera},\ and\ \citenamefont
  {Wu}}]{Wang2016}%
  \BibitemOpen
  \bibfield  {author} {\bibinfo {author} {\bibfnamefont {W.}~\bibnamefont
  {Wang}}, \bibinfo {author} {\bibfnamefont {C.-Z.}\ \bibnamefont {Chang}},
  \bibinfo {author} {\bibfnamefont {J.~S.}\ \bibnamefont {Moodera}}, \ and\
  \bibinfo {author} {\bibfnamefont {W.}~\bibnamefont {Wu}},\ }\href
  {http://dx.doi.org/10.1038/npjquantmats.2016.23} {\bibfield  {journal}
  {\bibinfo  {journal} {Npj Quantum Materials}\ }\textbf {\bibinfo {volume}
  {1}},\ \bibinfo {pages} {16023 EP } (\bibinfo {year} {2016})}\BibitemShut
  {NoStop}%
\bibitem [{\citenamefont {Checkelsky}\ \emph {et~al.}(2012)\citenamefont
  {Checkelsky}, \citenamefont {Ye}, \citenamefont {Onose}, \citenamefont
  {Iwasa},\ and\ \citenamefont {Tokura}}]{checkelsky2012}%
  \BibitemOpen
  \bibfield  {author} {\bibinfo {author} {\bibfnamefont {J.~G.}\ \bibnamefont
  {Checkelsky}}, \bibinfo {author} {\bibfnamefont {J.}~\bibnamefont {Ye}},
  \bibinfo {author} {\bibfnamefont {Y.}~\bibnamefont {Onose}}, \bibinfo
  {author} {\bibfnamefont {Y.}~\bibnamefont {Iwasa}}, \ and\ \bibinfo {author}
  {\bibfnamefont {Y.}~\bibnamefont {Tokura}},\ }\href@noop {} {\bibfield
  {journal} {\bibinfo  {journal} {Nature Physics}\ }\textbf {\bibinfo {volume}
  {8}},\ \bibinfo {pages} {729} (\bibinfo {year} {2012})}\BibitemShut {NoStop}%
\bibitem [{\citenamefont {{Yasuda}}\ \emph {et~al.}(2017)\citenamefont
  {{Yasuda}}, \citenamefont {{Mogi}}, \citenamefont {{Yoshimi}}, \citenamefont
  {{Tsukazaki}}, \citenamefont {{Takahashi}}, \citenamefont {{Kawasaki}},
  \citenamefont {{Kagawa}},\ and\ \citenamefont {{Tokura}}}]{AQH_RIKEN}%
  \BibitemOpen
  \bibfield  {author} {\bibinfo {author} {\bibfnamefont {K.}~\bibnamefont
  {{Yasuda}}}, \bibinfo {author} {\bibfnamefont {M.}~\bibnamefont {{Mogi}}},
  \bibinfo {author} {\bibfnamefont {R.}~\bibnamefont {{Yoshimi}}}, \bibinfo
  {author} {\bibfnamefont {A.}~\bibnamefont {{Tsukazaki}}}, \bibinfo {author}
  {\bibfnamefont {K.~S.}\ \bibnamefont {{Takahashi}}}, \bibinfo {author}
  {\bibfnamefont {M.}~\bibnamefont {{Kawasaki}}}, \bibinfo {author}
  {\bibfnamefont {F.}~\bibnamefont {{Kagawa}}}, \ and\ \bibinfo {author}
  {\bibfnamefont {Y.}~\bibnamefont {{Tokura}}},\ }\href@noop {} {\bibfield
  {journal} {\bibinfo  {journal} {ArXiv e-prints}\ } (\bibinfo {year}
  {2017})},\ \Eprint {http://arxiv.org/abs/1707.09105} {arXiv:1707.09105
  [cond-mat.mes-hall]} \BibitemShut {NoStop}%
\bibitem [{\citenamefont {Rosen}\ \emph {et~al.}(2017)\citenamefont {Rosen},
  \citenamefont {Fox}, \citenamefont {Kou}, \citenamefont {Pan}, \citenamefont
  {Wang},\ and\ \citenamefont {Goldhaber-Gordon}}]{AQH_GHG}%
  \BibitemOpen
  \bibfield  {author} {\bibinfo {author} {\bibfnamefont {I.~T.}\ \bibnamefont
  {Rosen}}, \bibinfo {author} {\bibfnamefont {E.~J.}\ \bibnamefont {Fox}},
  \bibinfo {author} {\bibfnamefont {X.}~\bibnamefont {Kou}}, \bibinfo {author}
  {\bibfnamefont {L.}~\bibnamefont {Pan}}, \bibinfo {author} {\bibfnamefont
  {K.~L.}\ \bibnamefont {Wang}}, \ and\ \bibinfo {author} {\bibfnamefont
  {D.}~\bibnamefont {Goldhaber-Gordon}},\ }\href {\doibase
  10.1038/s41535-017-0073-0} {\bibfield  {journal} {\bibinfo  {journal} {npj
  Quantum Materials}\ }\textbf {\bibinfo {volume} {2}},\ \bibinfo {pages} {69}
  (\bibinfo {year} {2017})}\BibitemShut {NoStop}%
\bibitem [{\citenamefont {Dzero}\ \emph {et~al.}(2010)\citenamefont {Dzero},
  \citenamefont {Sun}, \citenamefont {Galitski},\ and\ \citenamefont
  {Coleman}}]{DzeroTKI}%
  \BibitemOpen
  \bibfield  {author} {\bibinfo {author} {\bibfnamefont {M.}~\bibnamefont
  {Dzero}}, \bibinfo {author} {\bibfnamefont {K.}~\bibnamefont {Sun}}, \bibinfo
  {author} {\bibfnamefont {V.}~\bibnamefont {Galitski}}, \ and\ \bibinfo
  {author} {\bibfnamefont {P.}~\bibnamefont {Coleman}},\ }\href {\doibase
  10.1103/PhysRevLett.104.106408} {\bibfield  {journal} {\bibinfo  {journal}
  {Phys. Rev. Lett.}\ }\textbf {\bibinfo {volume} {104}},\ \bibinfo {pages}
  {106408} (\bibinfo {year} {2010})}\BibitemShut {NoStop}%
\bibitem [{\citenamefont {Samm}\ \emph {et~al.}(2014)\citenamefont {Samm},
  \citenamefont {Gramich}, \citenamefont {Baumgartner}, \citenamefont {Weiss},\
  and\ \citenamefont {Schönenberger}}]{NegTMR}%
  \BibitemOpen
  \bibfield  {author} {\bibinfo {author} {\bibfnamefont {J.}~\bibnamefont
  {Samm}}, \bibinfo {author} {\bibfnamefont {J.}~\bibnamefont {Gramich}},
  \bibinfo {author} {\bibfnamefont {A.}~\bibnamefont {Baumgartner}}, \bibinfo
  {author} {\bibfnamefont {M.}~\bibnamefont {Weiss}}, \ and\ \bibinfo {author}
  {\bibfnamefont {C.}~\bibnamefont {Schönenberger}},\ }\href {\doibase
  10.1063/1.4874919} {\bibfield  {journal} {\bibinfo  {journal} {Journal of
  Applied Physics}\ }\textbf {\bibinfo {volume} {115}},\ \bibinfo {pages}
  {174309} (\bibinfo {year} {2014})}\BibitemShut {NoStop}%
\bibitem [{\citenamefont {Vlasov}\ \emph {et~al.}(2017)\citenamefont {Vlasov},
  \citenamefont {Guillemette}, \citenamefont {Gervais},\ and\ \citenamefont
  {Szkopek}}]{vlasov2017}%
  \BibitemOpen
  \bibfield  {author} {\bibinfo {author} {\bibfnamefont {A.}~\bibnamefont
  {Vlasov}}, \bibinfo {author} {\bibfnamefont {J.}~\bibnamefont {Guillemette}},
  \bibinfo {author} {\bibfnamefont {G.}~\bibnamefont {Gervais}}, \ and\
  \bibinfo {author} {\bibfnamefont {T.}~\bibnamefont {Szkopek}},\ }\href@noop
  {} {\bibfield  {journal} {\bibinfo  {journal} {arXiv:1706.00458}\ } (\bibinfo
  {year} {2017})}\BibitemShut {NoStop}%
\bibitem [{\citenamefont {Upadhyaya}\ and\ \citenamefont
  {Tserkovnyak}(2016)}]{AQHPiston}%
  \BibitemOpen
  \bibfield  {author} {\bibinfo {author} {\bibfnamefont {P.}~\bibnamefont
  {Upadhyaya}}\ and\ \bibinfo {author} {\bibfnamefont {Y.}~\bibnamefont
  {Tserkovnyak}},\ }\href {\doibase 10.1103/PhysRevB.94.020411} {\bibfield
  {journal} {\bibinfo  {journal} {Phys. Rev. B}\ }\textbf {\bibinfo {volume}
  {94}},\ \bibinfo {pages} {020411} (\bibinfo {year} {2016})}\BibitemShut
  {NoStop}%
\bibitem [{\citenamefont {Tserkovnyak}\ and\ \citenamefont
  {Loss}(2012)}]{LossSpinTorque}%
  \BibitemOpen
  \bibfield  {author} {\bibinfo {author} {\bibfnamefont {Y.}~\bibnamefont
  {Tserkovnyak}}\ and\ \bibinfo {author} {\bibfnamefont {D.}~\bibnamefont
  {Loss}},\ }\href {\doibase 10.1103/PhysRevLett.108.187201} {\bibfield
  {journal} {\bibinfo  {journal} {Phys. Rev. Lett.}\ }\textbf {\bibinfo
  {volume} {108}},\ \bibinfo {pages} {187201} (\bibinfo {year}
  {2012})}\BibitemShut {NoStop}%
\bibitem [{\citenamefont {Garate}\ and\ \citenamefont
  {Franz}(2010)}]{GarateCorbino}%
  \BibitemOpen
  \bibfield  {author} {\bibinfo {author} {\bibfnamefont {I.}~\bibnamefont
  {Garate}}\ and\ \bibinfo {author} {\bibfnamefont {M.}~\bibnamefont {Franz}},\
  }\href {\doibase 10.1103/PhysRevLett.104.146802} {\bibfield  {journal}
  {\bibinfo  {journal} {Phys. Rev. Lett.}\ }\textbf {\bibinfo {volume} {104}},\
  \bibinfo {pages} {146802} (\bibinfo {year} {2010})}\BibitemShut {NoStop}%
\bibitem [{\citenamefont {Thouless}(1969)}]{Thouless1969}%
  \BibitemOpen
  \bibfield  {author} {\bibinfo {author} {\bibfnamefont {D.~J.}\ \bibnamefont
  {Thouless}},\ }\href@noop {} {\bibfield  {journal} {\bibinfo  {journal}
  {Physical Review}\ }\textbf {\bibinfo {volume} {187}},\ \bibinfo {pages}
  {732} (\bibinfo {year} {1969})}\BibitemShut {NoStop}%
\bibitem [{Note1()}]{Note1}%
  \BibitemOpen
  \bibinfo {note} {By assuming this relationship between $\gamma $ and $\gamma
  _-$, we simplify the solution to the rate equation. Moderate deviations in
  the annihilation rate $\gamma _-$ will not significantly influence the
  predictions made here for the shape of hysteresis curves resulting from $N\gg
  1$ spin-flip events.}\BibitemShut {Stop}%
\bibitem [{\citenamefont {Amasha}\ \emph {et~al.}(2008)\citenamefont {Amasha},
  \citenamefont {MacLean}, \citenamefont {Radu}, \citenamefont {Zumb\"uhl},
  \citenamefont {Kastner}, \citenamefont {Hanson},\ and\ \citenamefont
  {Gossard}}]{spinrelaxation}%
  \BibitemOpen
  \bibfield  {author} {\bibinfo {author} {\bibfnamefont {S.}~\bibnamefont
  {Amasha}}, \bibinfo {author} {\bibfnamefont {K.}~\bibnamefont {MacLean}},
  \bibinfo {author} {\bibfnamefont {I.~P.}\ \bibnamefont {Radu}}, \bibinfo
  {author} {\bibfnamefont {D.~M.}\ \bibnamefont {Zumb\"uhl}}, \bibinfo {author}
  {\bibfnamefont {M.~A.}\ \bibnamefont {Kastner}}, \bibinfo {author}
  {\bibfnamefont {M.~P.}\ \bibnamefont {Hanson}}, \ and\ \bibinfo {author}
  {\bibfnamefont {A.~C.}\ \bibnamefont {Gossard}},\ }\href {\doibase
  10.1103/PhysRevLett.100.046803} {\bibfield  {journal} {\bibinfo  {journal}
  {Phys. Rev. Lett.}\ }\textbf {\bibinfo {volume} {100}},\ \bibinfo {pages}
  {046803} (\bibinfo {year} {2008})}\BibitemShut {NoStop}%
\bibitem [{\citenamefont {Qassemi}\ \emph {et~al.}(2009)\citenamefont
  {Qassemi}, \citenamefont {Coish},\ and\ \citenamefont
  {Wilhelm}}]{qassemi2009}%
  \BibitemOpen
  \bibfield  {author} {\bibinfo {author} {\bibfnamefont {F.}~\bibnamefont
  {Qassemi}}, \bibinfo {author} {\bibfnamefont {W.~A.}\ \bibnamefont {Coish}},
  \ and\ \bibinfo {author} {\bibfnamefont {F.~K.}\ \bibnamefont {Wilhelm}},\
  }\href@noop {} {\bibfield  {journal} {\bibinfo  {journal} {\prl}\ }\textbf
  {\bibinfo {volume} {102}},\ \bibinfo {pages} {176806} (\bibinfo {year}
  {2009})}\BibitemShut {NoStop}%
\end{thebibliography}%

\end{document}